\documentclass[11pt]{article}
\usepackage{amssymb,amsmath,amsthm,amscd,latexsym}
\usepackage{mathrsfs}
\usepackage{amsfonts}
\usepackage{amsmath}
\usepackage{amssymb}
\usepackage{amscd}
\usepackage{amsfonts,caption2}
\usepackage{amsmath,amsthm,amssymb}
\usepackage{extpfeil}
\usepackage{color,xcolor}
\usepackage{graphicx}

\newtheorem{Theorem}{Theorem}
\newtheorem{Corollary}{Corollary}
\newtheorem{Definition}{Definition}
\newtheorem{Example}{Example}
\newtheorem{Remark}{Remark}

\newtheorem{Lemma}{Lemma}

\newcommand{\Z}{\mathbb{Z}}
\makeatletter
\newcommand{\romanNum}[1]{\@roman{#1}}
\makeatother

\begin{document}
\title{New constructions of $q$-Ary 2-D Z-Complementary Array Pairs}

\author{Hui Zhang\thanks{School of Mathematics, Southwest Jiaotong University, Chengdu, 610031, China. Email:875009327@qq.com}, Cuiling Fan\thanks{School of Mathematics, Southwest Jiaotong University, Chengdu, 610031, China. Email: fcl@swjtu.edu.cn}, and Sihem Mesnager\thanks{Department of Mathematics, University of Paris VIII, 93526 Saint-Denis, with University Sorbonne Paris Cit\'{e}, LAGA,
UMR 7539, CNRS, 93430 Villetaneuse, and also T\'{e}l\'{e}com Paris,
91120 Palaiseau, France. Email: smesnager@univ-paris8.fr}}

\date{\today}
\maketitle

\begin{abstract}
This paper is devoted to sequences and focuses on designing new two-dimensional (2-D) Z-complementary array pairs (ZCAPs) by exploring two promising approaches. A ZCAP is a pair of 2-D arrays, whose 2-D autocorrelation sum gives zero value at all time shifts in a zone around the $(0,0)$ time shift, except the $(0,0)$ time shift.
The first approach investigated in this paper uses a one-dimensional (1-D) Z-complementary pair (ZCP), which is an extension of the 1-D Golay complementary pair (GCP) where the autocorrelations of constituent sequences are complementary within a zero correlation zone (ZCZ). The second approach involves directly generalized Boolean functions (which are important components with many applications, particularly in (symmetric) cryptography). Along with this paper, new construction of 2-D ZCAPs is proposed based on 1-D ZCP, and direct construction of 2-D ZCAPs is also offered directly by 2-D generalized Boolean functions. Compared to existing constructions based on generalized Boolean functions, our proposed construction covers all of them.
ZCZ sequences are a class of spreading sequences having ideal auto-correlation and cross-correlation in a zone around the origin. In recent years, they have been extensively studied due to their crucial applications, particularly in quasi-synchronous code division multiple access systems. Our proposed 2-D ZCAPs based on 2-D generalized Boolean functions have larger 2-D $\mathrm{ZCZ}_{\mathrm{ratio}}=\frac{6}{7}$. Compared to the construction based on ZCPs, our proposed 2-D ZCAPs also have the largest 2-D $\mathrm{ZCZ}_{\mathrm{ratio}}$.\\

{\bf Keywords}: Generalized Boolean function \and Golay  complementary pair \and Z-complementary array pair \and Z-complementary pair \and Zero correlation zone.
\end{abstract}


\section{Introduction}
\label{intro}
In 1951, the concept of Golay  complementary pairs (GCPs) was
first proposed by M.J. Golay \cite{1951-Golay}. A pair of sequences is said to be an GCP if their aperiodic autocorrelation sums are zero except at zero shift. GCPs have found many engineering applications for its ideal correlation properties, such as inter-symbol interference channel estimation \cite{2001-Spasojevic}, radar wave from designs \cite{2018-Kumari}, and peak power control in orthogonal frequency division multiplexing (OFDM) \cite{1999-GDJ},\cite{2014-Wang}.
However, the length of $\mathrm{GCP}$ is very limited,
for example, the binary GCPs are known to exist for lengths of
the form $2^{\alpha}10^{\beta}26^{\gamma}$ where $\alpha$, $\beta$ and $\gamma$ are nonnegative integers \cite{2003-7-Borwein}. Hence, The Z-complementary pair (ZCP)  was proposed in \cite{2007-8-Fan} by introducing the concept of  zero correlation zone (ZCZ) to the GCP in 2007. The sum of autocorrelations of constituent sequences is zero within a range of shifts. According to
the relaxation of the autocorrelation constraint, ZCPs was
shown to exist for all lengths \cite{2011-1-Li}.

The 1-D ZCP has been extended to 2-D arrays called the Z-complementary array pair (ZCAP). Also, the 2-D ZCAP includes
the 2-D Golay complementary
array pair (GCAP) as a special case.
Likewise, 2-D GCAPs also have a good autocorrelation property. The
aperiodic autocorrelations of two arrays in a 2-D GCAP sum
up to zero except for the 2-D zero shift. 2-D GCAPs have found many engineering applications for their ideal correlation properties. For example, they can be applied in 2-D synchronization \cite{1982-Golomb},\cite{1983-Hershey}, radar \cite{1983-Weathers}, and can be used as spreading sequences in the 2-D multi-carrier code division
multiple access (MC-CDMA) system \cite{2004-Turcs},\cite{2003-Farka?}.
Similarly, the array sizes of 2-D GCAPs are also very limited.
For 2-D binary GCAPs, the known size of each dimension is also limited in the form of
$2^{\alpha}10^{\beta}26^{\gamma}$. Similar to the 1-D case, the 2-D ZCAP is proposed to have more flexible sizes.

Nowadays, there are many known constructions of 1-D ZCPs (see. e.g. \cite{2018-Adhikary},\cite{2017-Chen},\cite{1999-GDJ},\cite{2019-Gu},\cite{2014-Liu-even},\cite{2014-odd},\cite{2019-Shen},\cite{2018-Xie},\cite{2021-YU}). In contrast, there are few constructions of 2-D ZCAPs. In \cite{2004-5-zeng}, the concept of 2-D ZCAPs was studied by Zeng \emph{et al}.
Later in \cite{2011-Li}, periodic 2-D ZCAPs were considered, and 2-D ZCAPs were constructed by interleaving existing 2-D GCAPs. In \cite{2019-pai} and \cite{2021-pai}, 2-D ZCAPs can be obtained from existing 1-D ZCPs or 2-D ZCAPs via using methods of concatenation or Kronecker product.
In \cite{2020-pai},\cite{2021-4-pai},\cite{2021-Roy}, 2-D ZCAPs based on 2-D generalized Boolean functions have been proposed.
In this paper, a new construction of 2-D ZCAPs is proposed based on 1-D ZCPs, and direct construction of 2-D ZCAPs also has been proposed based on 2-D generalized Boolean functions. The construction based on  1-D ZCPs including the construction of \cite{2020-pai},\cite{2021-4-pai}, and compared to the construction of \cite{2019-pai},\cite{2021-pai}, our proposed 2-D ZCAPs also have the largest 2-D $\mathrm{ZCZ}_{\mathrm{ratio}}$. Moreover, compared to \cite{2021-4-pai},\cite{2021-Roy}, our proposed 2-D ZCAPs based on 2-D generalized Boolean functions have the largest 2-D $\mathrm{ZCZ}_{\mathrm{ratio}}=\frac{6}{7}$.

This article is organized as follows. In Section \ref{Sec-Preliminaries}, we give some basic notation and definitions of 1-D ZCP, 2-D ZCAP, and generalized Boolean functions. Section \ref{Sec-constructions} is the core of the paper in which we present our two approaches for designing 2-D ZCAPs. The first one (Subsection \ref{approach1}) uses 1-D ZCPs while the second one (Subsection \ref{approach2}) employees 2-D generalized Boolean functions.  Section \ref{Sec-Conclusions} concludes this paper after a comparison of our results with former constructions provided in the literature.

\section{Preliminaries}\label{Sec-Preliminaries}
This section recalls some definitions of 1-D ZCP, 2-D ZCAP, and generalized Boolean functions. Before then, we introduce the notations which will be used throughout the paper.
\begin{itemize}
  \item For positive integer $q$, $\Z_q=\{0,1,\cdots,q-1\}$.
  \item $\xi=e^{\frac{2\pi\sqrt{-1}}{q}}$ is a $q$th primitive root of unity.
  \item $(\cdot)^{*}$ denotes the complex conjugation.
  \item $(\cdot)^{T}$ denotes the transpose.
 \item A complex valued sequence $\mathbf{A}={\xi}^{\mathbf{a}}$, where $\mathbf{a}=(a_0,a_1,\cdots,a_{L-1})$, $a_i\in\Z_q$, $0\le i<L$.
\item A complex valued array $\mathcal{C}=\xi^{\mathbf{\mathcal{c}}}$, where $\mathcal{C}=(C_{i,g})$ and $\mathbf{\mathcal{c}}=(c_{i,g})$, $c_{i,g}\in\Z_q$, $0\le i<L_1$, $0\le g<L_2$.
  \item $\overleftarrow{{\mathbf{a}}}=(a_{L-1},a_{L-2},\cdots,a_0)$.
\end{itemize}

In this paper, we will consistently use lower-case boldface letters for sequences over $\Z_q$ and upper-case boldface letters for complex-valued sequences; the same letter (for example $\mathbf{a}$ and $\mathbf{A}$) will indicate that the sequences correspond. Similarly, we will use lower-case cursive letters for arrays over $\Z_q$ and upper-case cursive letters for complex-valued arrays; the same letter (for example $\mathcal{c}$ and $\mathcal{C}$) will indicate that the arrays correspond.
\begin{Definition}
For two complex valued sequences $\mathbf{A}=(A_0,A_1,\cdots,A_{L-1})$ and $\mathbf{B}=(B_0,B_1,\cdots,B_{L-1})$ of
length $L$, the aperiodic cross-correlation function $(\mathrm{ACCF})$ for
time shift $u$ is defined by
\begin{equation*}
 \rho(\mathbf{A},\mathbf{B};u)=
\left\{
\begin{array}{ll}
\sum\limits_{i=0}^{L-1-u}A_{i+u}B^{*}_{i},&0\le u<L;\\
\sum\limits_{i=0}^{L-1+u}A_{i}B^{*}_{i-u},&-L<u<0;\\
0,& |u|\ge L;\\
\end{array}
\right.
\end{equation*}
where $(\cdot)^{*}$ denotes the complex conjugate. If $\mathbf{A}=\mathbf{B}$, then the
function is called aperiodic autocorrelation function $(\mathrm{AACF})$ and is denoted by $\rho(\mathbf{A};u)$.
\end{Definition}
\begin{Definition}
A pair of sequences $\mathbf{A}$ and $\mathbf{B}$ of length
$L$ is a $\mathrm{ZCP}$, denoted by $(L,Z)-\mathrm{ZCP}$, if and only if
\begin{equation*}
 \rho(\mathbf{A};u)+\rho(\mathbf{B};u)=
\left\{
\begin{array}{ll}
2L,&u=0;\\
0,& -Z<u<Z, u\neq 0;\\
\end{array}
\right.
\end{equation*}
where $Z\le L$. Here $Z$ is called the $\mathrm{ZCZ}$ width.
If $Z=L$, then $(\mathbf{A},\mathbf{B})$ is called a $\mathrm{GCP}$.
\end{Definition}
\begin{Definition}
Let $(\mathbf{A},\mathbf{B})$ and $(\mathbf{A}_1,\mathbf{B}_1)$ be two $\mathrm{ZCPs}$ of length $L$ and $\mathrm{ZCZ}$ width $Z$. Then they are said to be Z-complementary mate to each other if they satisfy the following
\begin{equation*}
 \rho(\mathbf{A},\mathbf{A}_1;u)+\rho(\mathbf{B},\mathbf{B}_1;u)=0,~\mbox{for}~|u|<Z.
\end{equation*}
When $Z=N$, then $(\mathbf{A},\mathbf{B})$ and $(\mathbf{A}_1,\mathbf{B}_1)$  are called complementary mates to each other.
\end{Definition}
\subsection{Two Dimensional Arrays}
\begin{Definition}
$2$-D $\mathrm{ACCF}$ of two arrays $\mathcal{C}$ and $\mathcal{D}$ at shift $(u_1,u_2)$ is defined as
\begin{eqnarray*}
\rho(\mathcal{C},\mathcal{D};u_1,u_2)=
\left\{
\begin{array}{ll}
\sum\limits_{i=0}^{L_1-1-u_1}\sum\limits_{g=0}^{L_2-1-u_2}{{D_{{i+u_1},{g+u_2}}}D_{i,g}^{*}},& 0\le u_1<L_1,0\le u_2<L_2;\\
\sum\limits_{i=0}^{L_1-1-u_1}\sum\limits_{g=0}^{L_2-1+u_2}{{D_{i+u_1,g}}D_{i,{g-u_2}}^{*}},& 0\le u_1<L_1,-L_2<u_2<0;\\
\sum\limits_{i=0}^{L_1-1+u_1}\sum\limits_{g=0}^{L_2-1-u_2}{{D_{{i},{g+u_2}}}D_{{i-u_1},g}^{*}},& -L_1<u_1<0,0\le u_2<L_2;\\
\sum\limits_{i=0}^{L_1-1+u_1}\sum\limits_{g=0}^{L_2-1+u_2}{{D_{i,g}}D_{{i-u_1},{g-u_2}}^{*}},& -L_1<u_1<0,-L_2<u_2<0;\\
0,&\mbox{otherwise};
\end{array}
\right.
\end{eqnarray*}
\end{Definition}
\begin{Definition}
A pair of arrays $\mathcal{C}$ and $\mathcal{D}$ of size $L_1\times L_2$ is called an $((L_1,L_2)$,$(Z_1,Z_2))-\mathrm{ZCAP}$, if
\begin{eqnarray*}
\rho(\mathcal{C};u_1,u_2)+\rho(\mathcal{D};u_1,u_2)
=
\left\{
\begin{array}{lll}
2L_1L_2,& (u_1,u_2)=(0,0);\\
0,&0\le |u_1|<Z_1, 0\le |u_2|<Z_2, (u_1,u_2)\neq (0,0);\\
\end{array}
\right.
\end{eqnarray*}
where $Z_1\times Z_2$ is the size of the rectangular $\mathrm{ZCZ}$. If $\mathcal{C}=(\xi^{c_{i,g}})$ and $\mathcal{D}=(\xi^{d_{i,g}})$ where $\mathcal{c}=(c_{i,g})$ and $\mathcal{d}=(d_{i,g})$ over $\Z_q$ for $0\le i<L_1$, $0\le g<L_2$, then this array pair $(\mathcal{c},\mathcal{d})$ is also called a $q$-ary $\mathrm{ZCAPs}$.
\end{Definition}
\begin{Definition}
The $2$-D $\mathrm{ZCZ}$ ratio of an
$((L_1, L_2),(Z_1, Z_2))-\mathrm{ZCAP}$ as the ratio of the rectangular $\mathrm{ZCZ}$ over the array size given by
\begin{equation*}
\begin{array}{c}
\mathrm{ZCZ}_{\mathrm{ratio}}=\frac{Z_1Z_2}{L_1L_2}.
\end{array}
\end{equation*}
Note that when $\mathrm{ZCZ}_{\mathrm{ratio}}$ achieves its maximum value $1$. If $Z_1 = L_1$ and $Z_2 = L_2$, this array pair is a $2$-D $\mathrm{GCAP}$.
\end{Definition}
\subsection{Generalized Boolean Functions}
A generalized Boolean function is a function $f:\Z_2^{m}\longrightarrow\Z_q$ consisting of $m$ variables $x_1,x_2,\cdots, x_m$, where $x_l\in\Z_2$ for $l =1,2,\cdots,m$. A generalized Boolean function can be written uniquely as a $\Z_q$-valued function of the $2^m$ monomials $\{1,x_1,x_2,\cdots,x_{m},x_1x_2,\cdots,x_1x_2x_3,\cdots,x_1x_2\cdots x_{m}\}$, called algebraic normal form. For a $q$-ary generalized Boolean function with $m$ variables, we define the associated sequence $\mathbf{f}=(f_0,f_1,\cdots,f_{2^m-1})$ and let $f_i=f(i_1,i_2,\cdots,i_m)$ where $(i_1,i_2,\cdots,i_m)$ is the binary representation vector of the integer $i=\sum\limits_{l=1}^{m}i_l2^{m-l}$.

The length of a sequence $\mathbf{f}$ constructed by the Boolean function with $m$ variable is $2^m$. But, sometimes, we need sequence lengths that are not limited to be powers of $2$.
Therefore, we define the truncated sequence $\mathbf{f}^{L}$, where $\mathbf{f}^{L}$ is the result by removing the last $2^m-L$ elements from $\mathbf{f}$, that is, $\mathbf{f}^{L}=(f_0,f_1,\cdots,f_{L-1})$. Note that $f_i=f(i_1,i_2,\cdots,i_m)$, where $i=0,1,\cdots,L-1$.  What's more, we can further associate sequence $\mathbf{f}^{L}$ with a complex-valued sequence $\mathbf{F}^{L}$, and define another corresponding truncated complex-valued sequence ${\mathbf{F}}^{L}=\xi^{{\mathbf{f}}^{L}}$.

A 2-D generalized Boolean function is a function
$f: \Z_2^{n+m}\longrightarrow\Z_q$ consisting of $n+m$ variables
$x_1,x_2,\cdots,x_n,y_1,y_2,\cdots,y_m$, where $x_s, y_l\in\Z_2$ for $s =1,2,\cdots,n$ and $l=1,2,\cdots,m$ (\cite{2020-pai}).
For a 2-D generalized Boolean function $f$ of $n+m$
variables, we specify an array
\begin{equation*}
(f_{i,g})=
\left(
  \begin{array}{cccc}
    f_{0,0} & f_{0,1}  & \cdots & f_{0,2^m-1}  \\
    f_{1,0} & f_{1,1}  & \cdots & f_{1,2^m-1}  \\
    \vdots & \vdots & \ddots & \vdots \\
    f_{2^n-1,0} & f_{2^n-1,1}  & \cdots & f_{2^n-1,2^m-1}  \\
  \end{array}
\right)
\end{equation*}
of size $2^n\times 2^m$ by letting $f_{i,g}=f((i_1,i_2,\cdots,i_{n}),(g_1,g_2,\cdots,g_{m}))$ where $(i_1,i_2,\cdots,i_{n})$
and $(g_1,g_2,\cdots,g_{m})$ are binary representations of the integers
$i=\sum\limits_{s=1}^{n}i_s2^{n-s}$, $g=\sum\limits_{l=1}^{m}g_l2^{m-l}$, respectively.
Similar 1-D generalized Boolean function, we can associate array $\mathcal{f}=(f_{i,g})_{0\le i<2^n,0\le g<2^m}$ with a complex-valued array of size $2^n\times 2^m$ $\mathcal{F}=\xi^{\mathcal{f}}$, and define an other corresponding truncated complex-valued array $\mathcal{F}^{L_1\times L_2}=\xi^{(f_{i,g})_{0\le i<L_1,0\le g<L_2}}$.
\begin{Example}
Let $q=2$, $n=2$ and $m=3$, the associated array to the 2-D Boolean function $f=x_1x_2+x_1y_1+y_3$ is given by
\begin{equation*}
\mathcal{f}=
\left(
  \begin{array}{cccccccc}
    0 & 1 & 0 & 1 & 0 & 1 & 0 & 1 \\
    0 & 1 & 0 & 1 & 0 & 1 & 0 & 1 \\
    0 & 1 & 0 & 1 & 1 & 0 & 1 & 0 \\
    1 & 0 & 1 & 0 & 0 & 1 & 0 & 1 \\
  \end{array}
\right).
\end{equation*}
\end{Example}
\begin{Lemma}(\cite{1999-GDJ})
Let $\mathbf{x}=(x_1,x_2,\cdots,x_m)\in\Z_2^m$ and
\begin{equation*}
f=\frac{q}{2}\sum\limits_{l=1}^{m-1}x_{\pi(l)}x_{\pi(l+1)}+\sum\limits_{l=1}^{m}v_lx_l+v_0,
\end{equation*}
where $\pi$ is a permutation of the set $\{1,2,\cdots,m\}$, and $v_l\in\Z_q$, $0\le l\le m$.
Then the sequence pair $(\mathbf{a},\mathbf{b})$ given by
\begin{equation*}
\left\{
\begin{array}{l}
\mathbf{a}=\mathbf{f}\\
\mathbf{b}=\mathbf{f}+\frac{q}{2}\mathbf{x}_{\pi(1)}
\end{array}
\right.
\end{equation*}
is a $\mathrm{GCP}$ over $\Z_q$ of length $2^m$.
What's more, the sequence pair
\begin{equation*}
(\mathbf{f},\mathbf{f}+\frac{q}{2}\mathbf{x}_{\pi(m)})
\end{equation*}
is a $\mathrm{GCP}$ of length $2^m$ over $\Z_q$.
\end{Lemma}
\section{Construction approaches for designing 2-D ZCAPs}\label{Sec-constructions}
The ultimate objective of this section is to design 2-D ZCAPs. We shall follow two directions as presented in the next subsections  \ref{approach1} and  \ref{approach2}.

\subsection{Our first approach for constructing 2-D ZCAPs}\label{approach1}
\begin{Lemma}(\cite{2002-Craigena})\label{H_2}
Let $(\mathbf{C},\mathbf{D})$ be a $\mathrm{GCP}$. Then $\rho(\overleftarrow{\mathbf{D}}^{*};u)+\rho(\overleftarrow{\mathbf{C}}^{*};u)=0$ and $\rho(\overleftarrow{\mathbf{D}};u)+\rho(\overleftarrow{\mathbf{C}};u)=0$ for all $0<u<L$.
\end{Lemma}
\begin{Lemma}\label{H_1}
Let $(\mathbf{C},\mathbf{D})$ be a $\mathrm{ZCP}$ of length $L$ and $\mathrm{ZCZ}$ width $Z$. Then, $(\overleftarrow{\mathbf{D}}^{*},-\overleftarrow{\mathbf{C}}^{*})$ is also a $\mathrm{ZCP}$ of the same length $L$ and the same $\mathrm{ZCZ}$ width, and
\begin{eqnarray*}
\rho(\mathbf{C},\overleftarrow{\mathbf{D}}^{*};u)+\rho(\mathbf{D},-\overleftarrow{\mathbf{C}}^{*};u)=0,
\end{eqnarray*}
where $0\le |u|<Z$.
\end{Lemma}
\begin{Lemma}\label{L_Yu}(\cite{2021-YU})
Let $(\mathbf{A},\mathbf{B})$ be a $\mathrm{GCP}$ of length $L$, and $(\mathbf{C},\mathbf{D}) =
(\overleftarrow{\mathbf{B}}^{*},-\overleftarrow{\mathbf{A}}^{*})$ be one of its mate. Then
\begin{equation*}
 \left\{
 \begin{array}{ll}
  \mathbf{S}=(\mathbf{A}||\mathbf{C}||\mathbf{A}||\mathbf{A}||-\mathbf{A}||\mathbf{A}||-\mathbf{C}||\mathbf{A}||-\mathbf{C}||-\mathbf{C}||\mathbf{C}||\mathbf{C}||-\mathbf{A}||-\mathbf{C})\\
  \mathbf{T}=(\mathbf{B}||\mathbf{D}||\mathbf{B}||\mathbf{B}||-\mathbf{B}||\mathbf{B}||-\mathbf{D}||\mathbf{B}||-\mathbf{D}||-\mathbf{D}||\mathbf{D}||\mathbf{D}||-\mathbf{B}||-\mathbf{D})
 \end{array}
  \right.
 \end{equation*}
is a $(14L,12L)-\mathrm{ZCP}$.
\end{Lemma}
Below we present our first main result.
 \begin{Theorem}
Suppose $(\mathbf{A},\mathbf{B})$ is a binary $(L_1,Z_1)-\mathrm{ZCP}$ and $(\mathbf{C},\mathbf{D})$ is an $(L_2,Z_2)-\mathrm{ZCP}$, $N=L_1L_2$. Let
\begin{equation}\label{ST_1}\tag{1}
 \left\{
 \begin{array}{l}
  S_{i,g}=\frac{A_i+B_i}{2}C_g+\frac{A_i-B_i}{2}\overleftarrow{D}^{*}_g,\\
  T_{i,g}=\frac{A_i+B_i}{2}D_g-\frac{A_i-B_i}{2}\overleftarrow{C}^{*}_g.
 \end{array}
  \right.
 \end{equation}
Then, $(\mathcal{S},\mathcal{T})$ is an $((L_1,L_2),(Z_1,Z_2))-\mathrm{ZCAP}$.
In particular, If $L_1=Z_1$, $L_2=Z_2$, $(\mathcal{S},\mathcal{T})$ is an $(L_1,L_2)-\mathrm{GCAP}$.
\begin{proof}
Here, we assume $(u_1,u_2)\neq (0,0)$. If $0\le u_1<Z_1,0\le u_2<Z_2$, we can obtain that
\begin{eqnarray*}
&&S_{{i+u_1},{g+u_2}}S^{*}_{i,g}\\
&=&[\frac{A_{i+u_1}+B_{i+u_1}}{2}C_{g+u_2}+\frac{A_{i+u_1}-B_{i+u_1}}{2}\overleftarrow{D}^{*}_{g+u_2}][\frac{A_{i}+B_{i}}{2}C^{*}_{g}+\frac{A_{i}-B_{i}}{2}\overleftarrow{D}_{g}]\\
&=&\frac{1}{4}[(A_{i+u_1}A_{i}+A_{i+u_1}B_{i}+B_{i+u_1}A_{i}+B_{i+u_1}B_{i})C_{g+u_2}C^{*}_{g}\\
&~&+(A_{i+u_1}A_{i}-A_{i+u_1}B_{i}+B_{i+u_1}A_{i}-B_{i+u_1}B_{i})C_{g+u_2}\overleftarrow{D}_{g}\\
&~&+(A_{i+u_1}A_{i}+A_{i+u_1}B_{i}-B_{i+u_1}A_{i}-B_{i+u_1}B_{i})\overleftarrow{D}^{*}_{g+u_2}C^{*}_{g}\\
&~&+(A_{i+u_1}A_{i}-A_{i+u_1}B_{i}-B_{i+u_1}A_{i}+B_{i+u_1}B_{i})\overleftarrow{D}^{*}_{g+u_2}\overleftarrow{D}_{g}]
\end{eqnarray*}
and
\begin{eqnarray*}
&&T_{{i+u_1},{g+u_2}}T^{*}_{i,g}\\
&=&[\frac{A_{i+u_1}+B_{i+u_1}}{2}D_{g+u_2}-\frac{A_{i+u_1}-B_{i+u_1}}{2}\overleftarrow{C}^{*}_{g+u_2}][\frac{A_{i}+B_{i}}{2}D^{*}_{g}-\frac{A_{i}-B_{i}}{2}\overleftarrow{C}_{g}]\\
&=&\frac{1}{4}[(A_{i+u_1}A_{i}+A_{i+u_1}B_{i}+B_{i+u_1}A_{i}+B_{i+u_1}B_{i})D_{g+u_2}D^{*}_{g}\\
&~&+(-A_{i+u_1}A_{i}+A_{i+u_1}B_{i}-B_{i+u_1}A_{i}+B_{i+u_1}B_{i})D_{g+u_2}\overleftarrow{C}_{g}\\
&~&+(-A_{i+u_1}A_{i}-A_{i+u_1}B_{i}+B_{i+u_1}A_{i}+B_{i+u_1}B_{i})\overleftarrow{C}^{*}_{g+u_2}D^{*}_{g}\\
&~&+(A_{i+u_1}A_{i}-A_{i+u_1}B_{i}-B_{i+u_1}A_{i}+B_{i+u_1}B_{i})\overleftarrow{C}^{*}_{g+u_2}\overleftarrow{C}_{g}].
\end{eqnarray*}
Therefore, we have
\begin{eqnarray*}
&&S_{{i+u_1},{g+u_2}}S^{*}_{i,g}+T_{{i+u_1},{g+u_2}}T^{*}_{i,g}\\
&=&\frac{1}{4}[(A_{i+u_1}A_{i}+A_{i+u_1}B_{i}+B_{i+u_1}A_{i}+B_{i+u_1}B_{i})(C_{g+u_2}C^{*}_{g}+D_{g+u_2}D^{*}_{g})\\
&~&+(A_{i+u_1}A_{i}-A_{i+u_1}B_{i}+B_{i+u_1}A_{i}-B_{i+u_1}B_{i})(C_{g+u_2}\overleftarrow{D}_{g}-D_{g+u_2}\overleftarrow{C}_{g})\\
&~&+(A_{i+u_1}A_{i}+A_{i+u_1}B_{i}-B_{i+u_1}A_{i}-B_{i+u_1}B_{i})(\overleftarrow{D}^{*}_{g+u_2}C^{*}_{g}-\overleftarrow{C}^{*}_{g+u_2}D^{*}_{g})\\
&~&+(A_{i+u_1}A_{i}-A_{i+u_1}B_{i}-B_{i+u_1}A_{i}+B_{i+u_1}B_{i})(\overleftarrow{D}^{*}_{g+u_2}\overleftarrow{D}_{g}+\overleftarrow{C}^{*}_{g+u_2}\overleftarrow{C}_{g})].
\end{eqnarray*}

When $0<u_1<Z_1$, $0\le u_2<Z_2$, we have
\begin{eqnarray*}
&&\rho(\mathcal{S};u_1,u_2)+\rho(\mathcal{T};u_1,u_2)\\
&=&\sum\limits_{i=0}^{L_1-1-u_1}\sum\limits_{g=0}^{L_2-1-u_2}(S_{{i+u_1},{g+u_2}}S^{*}_{i,g}+T_{{i+u_1},{g+u_2}}T^{*}_{i,g})\\
&=&\frac{1}{4}[(\rho(\mathbf{A};u_1)+\rho(\mathbf{A},\mathbf{B};u_1)+\rho(\mathbf{B},\mathbf{A};u_1)+\rho(\mathbf{B};u_1))(\rho(\mathbf{C};u_2)+\rho(\mathbf{D};u_2))\\
&~&+(\rho(\mathbf{A};u_1)-\rho(\mathbf{A},\mathbf{B};u_1)+\rho(\mathbf{B},\mathbf{A};u_1)-\rho(\mathbf{B};u_1))(\rho(\mathbf{C},\overleftarrow{\mathbf{D}}^{*};u_2)+\rho(\mathbf{D},-\overleftarrow{\mathbf{C}}^{*};u_2))\\
&~&+(\rho(\mathbf{A};u_1)+\rho(\mathbf{A},\mathbf{B};u_1)-\rho(\mathbf{B},\mathbf{A};u_1)-\rho(\mathbf{B};u_1))(\rho(\overleftarrow{\mathbf{D}}^{*},\mathbf{C};u_2)+\rho(-\overleftarrow{\mathbf{C}^{*}},\mathbf{D};u_2))\\
&~&+(\rho(\mathbf{A};u_1)-\rho(\mathbf{A},\mathbf{B};u_1)-\rho(\mathbf{B},\mathbf{A};u_1)+\rho(\mathbf{B};u_1))(\rho(\overleftarrow{\mathbf{D}}^{*};u_2)+\rho(\overleftarrow{\mathbf{C}}^{*};u_2))]\\
&=&0,
\end{eqnarray*}
where $2(\rho(\mathbf{C};u_2)+\rho(\mathbf{D};u_2))(\rho(\mathbf{A};u_1)+\rho(\mathbf{B};u_1))=0$ holds since $(\mathbf{A},\mathbf{B})$ is a $\mathrm{ZCP}$ and Lemma \ref{H_2}. By Lemma \ref{H_1}, $\rho(\mathbf{C},\overleftarrow{\mathbf{D}}^{*};u_2)+\rho(\mathbf{D},-\overleftarrow{\mathbf{C}}^{*};u_2)=0$, $\rho(\overleftarrow{\mathbf{D}}^{*},\mathbf{C};u_2)+\rho(-\overleftarrow{\mathbf{C}^{*}},\mathbf{D};u_2)=0$.

When $u_1=0$, $0<u_2<Z_2$, we have
\begin{eqnarray*}
&&\rho(\mathcal{S};u_1,u_2)+\rho(\mathcal{T};u_1,u_2)\\
&=&\sum\limits_{i=0}^{L_1-1}\sum\limits_{g=0}^{L_2-1-u_2}(S_{{i+u_1},{g+u_2}}S^{*}_{i,g}+T_{{i+u_1},{g+u_2}}T^{*}_{i,g})\\
&=&\frac{1}{4}[(\rho(\mathbf{A};0)+\rho(\mathbf{A},\mathbf{B};0)+\rho(\mathbf{B},\mathbf{A};0)+\rho(\mathbf{B};0))(\rho(\mathbf{C};u_2)+\rho(\mathbf{D};u_2))\\
&~&+(\rho(\mathbf{A};0)-\rho(\mathbf{A},\mathbf{B};0)+\rho(\mathbf{B},\mathbf{A};0)-\rho(\mathbf{B};0))(\rho(\mathbf{C},\overleftarrow{\mathbf{D}}^{*};u_2)-\rho(\mathbf{D},\overleftarrow{\mathbf{C}}^{*};u_2))\\
&~&+(\rho(\mathbf{A};0)+\rho(\mathbf{A},\mathbf{B};0)-\rho(\mathbf{B},\mathbf{A};0)-\rho(\mathbf{B};0))(\rho(\overleftarrow{\mathbf{D}}^{*},\mathbf{C};u_2)-\rho(\overleftarrow{\mathbf{C}^{*}},\mathbf{D};u_2))\\
&~&+(\rho(\mathbf{A};0)-\rho(\mathbf{A},\mathbf{B};0)-\rho(\mathbf{B},\mathbf{A};0)+\rho(\mathbf{B};0))(\rho(\overleftarrow{\mathbf{D}}^{*};u_2)+\rho(\overleftarrow{\mathbf{C}}^{*};u_2))]\\
&=&0.
\end{eqnarray*}

Similarly, we have
\begin{eqnarray*}
\rho(\mathcal{S};u_1,u_2)+\rho(\mathcal{T};u_1,u_2)=0,&0<|u_1|<Z_1,0<|u_2|<Z_2, (u_1,u_2)\neq (0,0).
\end{eqnarray*}

Hence,
 \begin{eqnarray*}
\rho(\mathcal{S};u_1,u_2)+\rho(\mathcal{T};u_1,u_2)
=
\left\{
\begin{array}{lll}
2L_1L_2,& (u_1,u_2)=(0,0);\\
0,&0\le |u_1|<Z_1, 0\le |u_2|<Z_2, (u_1,u_2)\neq (0,0).\\
\end{array}
\right.
\end{eqnarray*}
\end{proof}
 \end{Theorem}
\begin{Example}\label{E_2}
Suppose $(\mathbf{A},\mathbf{B})$ is a binary $(12,8)-\mathrm{ZCP}$ and $(\mathbf{C},\mathbf{D})$ is a $(4,4)-\mathrm{GCP}$. Let
\begin{equation*}
 \begin{array}{l}
\mathbf{A}=(-1,1,-1,-1,1,-1,1,1,1,-1,1,1),\\
\mathbf{B}=(-1,1,1,1,1,-1,-1,-1,1,-1,-1,-1),\\
\mathbf{C}=(1,1,j,-j),\\
\mathbf{D}=(1,1,-j,j),
\end{array}
\end{equation*}
where $j=\sqrt{-1}$. By \eqref{ST_1}, we have
\begin{eqnarray*}
\mathcal{S}^{T}&=&
\left(
\begin{array}{rrrrrrrrrrrr}
               -1 & 1  & j  & j  & 1 & -1 & -j& -j& 1 & -1 &-j & -j \\
               -1 & 1  &-j  & -j & 1 &-1  & j & j & 1 & -1 & j & j \\
               -j  & j  & -1 & -1 & j & -j & 1 & 1 & j & -j & 1 & 1 \\
               j &-j  & -1 &-1  &-j & j  & 1 & 1 & -j& j  & 1 & 1
\end{array}
\right),\\
\mathcal{T}^{T}&=&
\left(
  \begin{array}{rrrrrrrrrrrr}
               -1 & 1  & j  & j  & 1 & -1 & -j& -j & 1 & -1 & -j & -j \\
               -1 & 1  & -j & -j & 1 &-1  & j & j  & 1 & -1 & j  & j \\
                 j& -j & 1  & 1  & -j & j & -1& -1 & -j& j  & -1 & -1 \\
                -j& j  & 1  & 1  & j & -j & -1& -1 & j & -j &-1  & -1
\end{array}
\right),
\end{eqnarray*}
where $\mathcal{S}^{T}$ and $\mathcal{T}^{T}$ are the transpose of $\mathcal{S}$ and $\mathcal{T}$. The aperiodic autocorrelation sum is depicted in Fig.1.
\begin{figure}[htb]\label{Au_1}
\centering
\includegraphics[width=8cm]{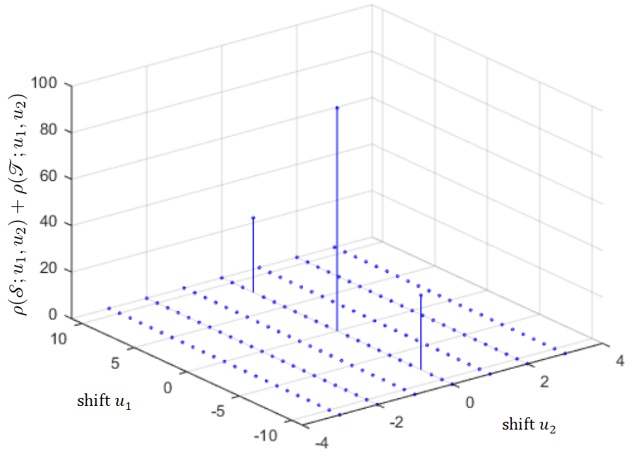}
\caption{Sum of autocorrelations of the $\mathrm{ZCAP}$ $(\mathcal{S},\mathcal{T})$ in Example \ref{E_2}.}
\end{figure}
Then, $(\mathcal{S},\mathcal{T})$ is a $((12,4),(8,4))-\mathrm{ZCAP}$.
\end{Example}
\begin{Corollary}\label{C_1}
 Suppose $({\mathbf{a}},{\mathbf{b}})$ is a binary $(L_1,Z_1)-\mathrm{ZCP}$ and $({\mathbf{c}},{\mathbf{d}})$ is a $q$-ary $(L_2,Z_2)-\mathrm{ZCP}$, $N=L_1L_2$. Define
 \begin{equation}\label{ST_2}\tag{2}
 \left\{
 \begin{array}{c}
  s_{i,g}=c_g-(\overleftarrow{d}_g{+}c_g)(a_i\oplus b_i)+\frac{q}{2}a_i,\\
  t_{i,g}=d_g-({\overleftarrow{d}_g{+}c_g})(a_i\oplus b_i)+\frac{q}{2}a_i,
 \end{array}
  \right.
 \end{equation}
where $\overleftarrow{\mathbf{d}}=(d_{L_{2}-1},d_{L_{2}-2},\cdots,d_{0})$, and $\oplus$ is addition modulo $2$.
Then, $(\mathcal{s},\mathcal{t})$ is a $q$-ary $((L_1,L_2),(Z_1,Z_2))-\mathrm{ZCAP}$.
In particular, If $L_1=Z_1$, $L_2=Z_2$, $(\mathcal{s},\mathcal{t})$ is a $q$-ary $(L_1,L_2)-\mathrm{GCAP}$.
\begin{proof}
It is easy to prove \eqref{ST_2} by \eqref{ST_1}.
\end{proof}
 \end{Corollary}
 \begin{Example}
Suppose $(\mathbf{a},\mathbf{b})$ is a binary $(12,8)-\mathrm{ZCP}$ and $(\mathbf{c},\mathbf{d})$ is a $(4,4)-\mathrm{GCP}$. Let
\begin{equation*}
 \begin{array}{l}
\mathbf{a}=(1,0,1,1,0,1,0,0,0,1,0,0),\\
\mathbf{b}=(1,0,0,0,0,1,1,1,0,1,1,1),\\
\mathbf{c}=(0,0,1,3),\\
\mathbf{d}=(0,0,3,1),
\end{array}
\end{equation*}
By \eqref{ST_2}, we have
\begin{eqnarray*}
\mathcal{s}^{T}&=&
\left(
\begin{array}{rrrrrrrrrrrr}
               2 & 0  & 1  & 1  & 0 & 2 & 3& 3& 0 & 2 &3 & 3 \\
               2 & 0  &3  & 3 & 0 &2  & 1 & 1 & 0 & 2 & 1 & 1 \\
               3  & 1  & 2 & 2 & 1 & 3 &0 & 0 & 1 & 3 & 0 & 0 \\
               1 &3  & 2 &2 &3 &1  & 0 & 0 & 3& 1  & 0 & 0
\end{array}
\right),\\
\mathcal{t}^{T}&=&
\left(
  \begin{array}{rrrrrrrrrrrr}
               2 & 0  & 1  & 1  & 0 & 2 & 3& 3 & 0 & 2 & 3 & 3 \\
               2 & 0  & 3 &3& 0 &2  & 1 & 1  & 0 &2 & 1  & 1 \\
                 1& 3& 0  & 0  & 3 & 1 &2& 2 & 3& 1  & 2 & 2 \\
                3& 1  & 0  &0  & 1 & 3& 2& 2 & 1 & 3 &2 & 2
\end{array}
\right),
\end{eqnarray*}
where $\mathcal{s}^{T}$ and $\mathcal{t}^{T}$ are the transpose of $\mathcal{s}$ and $\mathcal{t}$.
Then, $(\mathcal{s},\mathcal{t})$ is a quaternary $((12,4),(8,4))-\mathrm{ZCAP}$.
\end{Example}
It is  not difficult to see that given any two 1-D ZCPs,  one can construct a 2-D ZCAP.
\begin{Lemma}\label{ZCP_2}
Suppose $(\mathbf{c},\mathbf{d})$ is a $q$-ary $\mathrm{GDJ}$ of length $2^n$ and $(\mathbf{a},\mathbf{b})$ is a binary $(2^{m-1}+\sum\limits_{\alpha=t'+1}^{m-1}d_{\alpha}2^{\alpha-1}+2^v,2^{t'-1}+2^v)-\mathrm{ZCP}$, $v<t'<m$, $d_{\alpha}\in\Z_2$. Define
\begin{eqnarray*}
\left\{
\begin{array}{lll}
a_i=\sum\limits_{k=1}^{t'-1}i_{\pi_1(k)}i_{\pi_1(k+1)}+\sum\limits_{k=1}^{m}p_ki_k+p_0,\\
b_i=a_i+i_{{\pi_1}(1)},\\
c_g=\frac{q}{2}\sum\limits_{l=1}^{n-1}g_{{\pi_2(l)}}g_{{\pi_2(l+1)}}+\sum\limits_{l=1}^{n}v_lg_l+v_0,\\
d_g=c_g+\frac{q}{2}g_{\pi_2(1)}.
\end{array}
\right.
\end{eqnarray*}
Then we have
\begin{eqnarray*}
\left\{
\begin{array}{lll}
s_{g,i}=\frac{q}{2}\sum\limits_{l=1}^{n-1}g_{{\pi_2(l)}}g_{{\pi_2(l+1)}}+\sum\limits_{l=1}^{n}v'_lg_l+\frac{q}{2}g_{\pi_2(n)}i_{\pi_1(1)}+\frac{q}{2}\sum\limits_{k=1}^{t'-1}i_{\pi_1(k)}i_{\pi_1(k+1)}\\
~~\qquad+(\frac{q}{2}n-2v_0-\sum\limits_{l=1}^{n}v_l)i_{\pi_1(1)}+\sum\limits_{k=1}^{m}p'_ki_k+p'_0,\\
t_{g,i}=s_{i,g}+\frac{q}{2}g_{\pi_2(1)},
\end{array}
\right.
\end{eqnarray*}
where $v'_l=(1-2i_{\pi_1(1)})v_l$, $p'_k=\frac{q}{2}p_k$, $p'_0=\frac{q}{2}p_0$, $p_k\in\Z_2$, $v_l\in\Z_q$, $0\le k<m$, $0\le l<n$.
Hence, $(\mathcal{s},\mathcal{t})$ is a $q$-ary $((2^n,2^{m-1}+\sum\limits_{\alpha=t'+1}^{m-1}d_{\alpha}2^{\alpha-1}+2^v),(2^n,2^{t'-1}+2^v))-\mathrm{ZCAP}$.
\end{Lemma}
The previous Lemma follows straightforwardly from Corollary \ref{C_1}.
\begin{Remark}
We emphasize that when  $\frac{q}{2}n-2v_0-\sum\limits_{l=1}^{n}v_l\equiv0~(\mathrm{mod}~q)$ (as in Lemma \ref{ZCP_2}), we can recover the construction given by \cite{2021-4-pai} is a special case of Corollary  \ref{C_1}. Similarly, Theorem 6 in  \cite{2020-pai} is also a particular case of Corollary \ref{C_1}.

\end{Remark}
\subsection{Our second approach for constructing 2-D ZCAPs}\label{approach2}
In the following, we give a construction that has larger 2-D ZCZ ratios defined as the ratio of the ZCZ size over the array size by Corollary \ref{C_1}. Our construction is new compared with some of the existing ones.

\begin{Lemma}\label{L_6}
Suppose $(\mathbf{c},\mathbf{d})$ is a $q$-ary $\mathrm{GDJ}$ of length $2^m$ and $(\mathbf{a},\mathbf{b})$ is a binary $(14,12)-\mathrm{ZCP}$. Let
\begin{eqnarray*}
\left\{
\begin{array}{lll}
a_i=i_1+i_2+i_1i_2+i_1i_3+i_2i_4+i_1i_2i_4,\\
b_i=i_2+i_4+i_1i_3+i_2i_3+i_3i_4+i_1i_4+i_1i_2i_3+i_1i_2i_4+i_1i_3i_4,\\
c_g=\frac{q}{2}\sum\limits_{l=1}^{m-1}g_{{\pi(l)}}g_{{\pi(l+1)}}+\sum\limits_{l=1}^{m}v_lg_l+v_0,\\
d_g=c_g+\frac{q}{2}g_{\pi(1)},
\end{array}
\right.
\end{eqnarray*}
where $\pi$ be a permutation of $\{1,2,\cdots,m\}$, $v_l\in\Z_q$ and $0\le l\le m$.

Then we have
\begin{eqnarray*}
\left\{
\begin{array}{lll}
s_{i,g}&=&\frac{q}{2}\sum\limits_{l=1}^{m-1}g_{{\pi(l)}}g_{{\pi(l+1)}}+\sum\limits_{l=1}^{m}v_lg_l+v_0+\frac{q}{2}(i_1+i_2+i_1i_2+i_1i_3+i_2i_4+i_1i_2i_4)\\
&&+(\frac{q}{2}g_{\pi(m)}+\frac{q}{2}m-2\sum\limits_{l=1}^{m}v_lg_l-\sum\limits_{l=1}^{m}v_l-2v_0)(i_1+i_4+i_1i_2+i_2i_3+i_2i_4+i_3i_4\\ &&+i_1i_4+i_1i_2i_3+i_1i_3i_4),\\
t_{i,g}&=&s_{i,g}+\frac{q}{2}g_{\pi(1)},\\
\end{array}
\right.
\end{eqnarray*}
and the array pair $(\mathcal{s},\mathcal{t})$ is a q-$\mathrm{ary}$ $((14,2^m),(12,2^m))-\mathrm{ZCAP}$.
\end{Lemma}
We shall omit the proof of Lemma \ref{L_6} since it comes directly from Corollary \ref{C_1}. Now, thanks to Lemma \ref{L_6}, we provide in the following main result a direct construction of 2-D ZCAPs based on 2-D generalized Boolean functions.
\begin{Theorem}
Let $\mathbf{x}=(x_1,x_2,x_3,x_4,y_{1},y_{2},\cdots,y_{m})\in\Z_2^{4+m}$ and $\pi$ be a permutation of $\{1,2,\cdots,m\}$.
The $2$-D generalized Boolean function is given as
\begin{equation*}
\begin{array}{l}
f=\frac{q}{2}\sum\limits_{l=1}^{m-1}y_{{\pi(l)}}y_{{\pi(l+1)}}+\sum\limits_{l=1}^{m}v_ly_l+v_0+\frac{q}{2}(x_1+x_2+x_1x_2+x_1x_3+x_2x_4+x_1x_2x_4)\\
\qquad+(\frac{q}{2}m+\frac{q}{2}y_{\pi(m)}-2\sum\limits_{l=1}^{m}v_ly_l-\sum\limits_{l=1}^{m}v_l-2v_0)(x_1+x_4+x_1x_2+x_2x_3+x_2x_4+x_3x_4\\
\qquad+x_1x_4+x_1x_2x_3+x_1x_3x_4),\\
\end{array}
\end{equation*}
where $v_l\in\Z_q$, $0\le l\le m$ and $1\le k\le 4$,
the array pair
\begin{equation*}
(\mathbf{\mathcal{s}},\mathbf{\mathcal{t}})=(\mathcal{f},\mathcal{f}+\frac{q}{2}\mathcal{y}_{\pi(1)})
\end{equation*}
is a q-$\mathrm{ary}$ $((L_1,L_2),(Z_1,Z_2))-\mathrm{ZCAP}$, $N=L_1L_2=14\cdot 2^{m}$, $Z_1Z_2=12\cdot 2^{m}$, that is $\mathrm{ZCZ}_{\mathrm{ratio}}=\frac{6}{7}$ and $L_2=Z_2$.
Without loss of generality, set $L_1=14\cdot 2^{n}$, $L_2=2^{m-n}$, $Z_1=12\cdot 2^{n}$, $Z_2=2^{m-n}$, $0\le n\le m$. In particular, when $n=m$, the sequence pair $(\mathbf{s},\mathbf{t})$ is a q-$\mathrm{ary}$ $(14\cdot 2^{m},12\cdot 2^{m})-\mathrm{ZCP}$.
\begin{proof}
 For the array $\mathcal{s}$ of size $L_1\times L_2$, we let
\begin{equation*}
\mathcal{S}=\xi^{\mathcal{s}}=
\left(
  \begin{array}{cccc}
    \xi^{s_{0,0}} & \xi^{s_{0,1}} & \cdots & \xi^{s_{0,L_2-1}} \\
    \xi^{s_{1,0}} & \xi^{s_{2,1}} & \cdots & \xi^{s_{2,L_2-1}} \\
    \vdots& \vdots &\ddots&\vdots \\
\xi^{s_{L_1-1,0}} & \xi^{s_{L_1-1,1}} & \cdots&\xi^{s_{L_1-1,L_2-1}} \\
  \end{array}
\right),
\end{equation*}
where $\mathcal{s}$ can be expressed as
\begin{equation*}
\begin{array}{l}
s=\frac{q}{2}\sum\limits_{l=1}^{m-1}y_{{\pi(l)}}y_{{\pi(l+1)}}+\sum\limits_{l=1}^{m}v_ly_l+v_0+\frac{q}{2}(x_1+x_2+x_1x_2+x_1x_3+x_2x_4+x_1x_2x_4)\\
\qquad+(\frac{q}{2}m+\frac{q}{2}y_{\pi(m)}-2\sum\limits_{l=1}^{m}v_ly_l-\sum\limits_{l=1}^{m}v_l-2v_0)(x_1+x_4+x_1x_2+x_2x_3+x_2x_4+x_3x_4 \\
\qquad +x_1x_4+x_1x_2x_3+x_1x_3x_4).\\
\end{array}
\end{equation*}
Then, we need to prove that
\begin{equation}\label{S_1}\tag{3}
\begin{array}{l}
\rho(\mathcal{S};u_1,u_2)+\rho(\mathcal{T};u_1,u_2)\notag\\
=\sum\limits_{i=0}^{L_1-1-u_1}\sum\limits_{g=0}^{L_2-1-u_2}(\xi^{s_{{i+u_1},{g+u_2}}-s_{i,g}}+\xi^{t_{{i+u_1},{g+u_2}}-t_{i,g}})=0
\end{array}
\end{equation}
for $0\le u_1<Z_1$, $0\le u_2<Z_2$ and $(u_1,u_2)\neq (0,0)$.

For given $g$, $i$, we let $h=g+u_2$, $j=i+u_1$. We also let $(g_1,g_2,\cdots,g_{n})$, $(i_1,i_2,\cdots,i_{n})$, $(h_1,h_2,\cdots,h_{n})$ and $(j_1,j_2,\cdots,j_{n})$ be the binary representations of $g,i,h$ and $j$, respectively.
In what follows, we consider three cases to show that \eqref{S_1} holds.

\begin{itemize}
  \item[(1)] For $u_1>0$, $u_2\ge 0$, we suppose $g_{\pi(1)}\neq h_{\pi(1)}$. We can obtain
      \begin{equation*}
        s_{j,h}-s_{i,g}-t_{j,h}+t_{i,g}\equiv\frac{q}{2}(g_{\pi(1)}-h_{\pi(1)})\equiv\frac{q}{2}~(\mathrm{mod}~q)
      \end{equation*}
    implying $\xi^{s_{j,h}-s_{i,g}}/{\xi^{t_{j,h}-t_{i,g}}}=-1.$
      Therefore,
       \begin{equation*}
       \sum\limits_{i=0}^{L_1-1-u_1}(\xi^{s_{j,h}-s_{i,g}}+{\xi^{t_{j,h}-t_{i,g}}})=0.
      \end{equation*}
  \item[(2)] Suppose $u_1>0$, $u_2\ge 0$, $g_{\pi(1)}=h_{\pi(1)}$. Since $g\neq h$, we can define $k$ to be the smallest integer for which $g_{\pi(k)}\neq h_{\pi(k)}$. Let $g'$ and $h'$ be integers distinct from $g$ and $h$, respectively, only in one position $\pi(k-1)$. That is $g'_{\pi(k-1)}=1-g_{\pi(k-1)}$ and $h'_{\pi(k-1)}=1-h_{\pi(k-1)}$. Hence, we have
      \begin{eqnarray*}
      &&s_{i,g'}-s_{i,g}\\
       &=& \frac{q}{2}(g_{\pi(k-2)}g'_{\pi(k-1)}-g_{\pi(k-2)}g_{\pi(k-1)}+g'_{\pi(k-1)}g_{\pi(k)}-g_{\pi(k-1)}g_{\pi(k)})+v_{\pi(k-1)}g'_{\pi(k)}\\
&~&-v_{\pi(k-1)}g_{\pi(k)} \\
       &\equiv &\frac{q}{2}(g_{\pi(k-2)}+g_{\pi(k)})+v_{\pi(k-1)}(1-2g_{\pi(k-1)})~(\mathrm{mod}~q).
      \end{eqnarray*}
      Since $g_{\pi(k-2)}=h_{\pi(k-2)}$, $g_{\pi(k-1)}=h_{\pi(k-1)}$, we have
      \begin{eqnarray*}
      &&s_{j,h}-s_{i,g}-s_{j,h'}+s_{i,g'}\\
       &=& \frac{q}{2}(g_{\pi(k-2)}-h_{\pi(k-2)}+g_{\pi(k)}-h_{\pi(k)})+v_{\pi(k-1)}(2h_{\pi(k-1)}-2g_{\pi(k-1)}) \\
       &\equiv &\frac{q}{2}(g_{\pi(k)}-h_{\pi(k)})~(\mathrm{mod}~q)\\
       &\equiv &\frac{q}{2}~(\mathrm{mod}~q),
      \end{eqnarray*}
      which results in
      \begin{equation*}
       \xi^{s_{j,h}-s_{i,g}}+{\xi^{s_{j,h'}-s_{i,g'}}}=0.
      \end{equation*}
      Similarly, we can also obtain
      \begin{equation*}
       \xi^{t_{j,h}-t_{i,g}}+{\xi^{t_{j,h'}-t_{i,g'}}}=0.
      \end{equation*}
      Therefore,
       \begin{equation*}
       \sum\limits_{i=0}^{L_1-1-u_1}(      \xi^{s_{j,h}-s_{i,g}}+{\xi^{s_{j,h'}-s_{i,g'}}}+\xi^{t_{j,h}-t_{i,g}}+{\xi^{t_{j,h'}-t_{i,g'}}})=0.
      \end{equation*}

\item[(3)] For $u_1=0$ and $u_2>0$.

When $n=0$, $L_1=14$, set $i=\sum\limits_{j=1}^{4}i_j2^{4-j}$, $i=0,1,\cdots,13$.

Since $N=14\cdot 2^m$, $\mathbf{S}^{N}$ can be regarded as concatenating 14 complementary sequences of length $2^m$.
      Let $s_i=s(i_1,i_2,i_3,i_4,y_1,y_2,\cdots,y_m)$, $t_i=t(i_1,i_2,i_3,i_4,y_1,y_2,\cdots,y_m)$.
      It is obviously that $S_i=\xi^{s(i_1,i_2,i_3,i_4,y_1,y_2,\cdots,y_m)}$, $T_i=\xi^{t(i_1,i_2,i_3,i_4,y_1,y_2,\cdots,y_m)}$.
      Then
      \begin{eqnarray*}
\left\{
      \begin{array}{lll}
        \mathbf{S}^{N} &=& (\mathbf{S}_0||\mathbf{S}_1||\mathbf{S}_2||\cdots||\mathbf{S}_{13}), \\
        \mathbf{T}^{N} &=& (\mathbf{T}_0||\mathbf{T}_1||\mathbf{T}_2||\cdots||\mathbf{T}_{13}).
\end{array}
\right.
      \end{eqnarray*}
\begin{itemize}

\item If $(i_1,i_2,i_3,i_4)=\{(0,0,0,0),(0,0,1,0),(0,0,1,1),(0,1,0,1),(0,1,1,1)\}$, we have
\begin{equation*}
\left\{
\begin{array}{l}
 s_i=\frac{q}{2}\sum\limits_{l=1}^{m-1}y_{{\pi(l)}}y_{{\pi(l+1)}}+\sum\limits_{l=1}^{m}v_ly_l+v_0,\\
 t_i=s_i+\frac{q}{2}y_{\pi(1)}.
\end{array}
\right.
\end{equation*}

\item If $(i_1,i_2,i_3,i_4)=\{(0,1,0,0),(1,1,0,0)\}$, we have
\begin{equation*}
\left\{
\begin{array}{l}
 s_i=\frac{q}{2}\sum\limits_{l=1}^{m-1}y_{{\pi(l)}}y_{{\pi(l+1)}}+\sum\limits_{l=1}^{m}v_ly_l+v_0+\frac{q}{2},\\
 t_i=s_i+\frac{q}{2}y_{\pi(1)}.
\end{array}
\right.
\end{equation*}

\item If $(i_1,i_2,i_3,i_4)=\{(0,0,0,1),(1,0,1,0),(1,0,1,1)\}$, we have
\begin{equation*}
\left\{
\begin{array}{l}
s_i=\frac{q}{2}\sum\limits_{l=1}^{m-1}y_{{\pi(l)}}y_{{\pi(l+1)}}-\sum\limits_{l=1}^{m}v_ly_l-v_0+(\frac{q}{2}m+\frac{q}{2}y_{\pi(m)}-\sum\limits_{l=1}^{n}v_l),\\
t_i=s_i+\frac{q}{2}y_{\pi(1)}.
\end{array}
\right.
\end{equation*}

\item If $(i_1,i_2,i_3,i_4)=\{(0,1,1,0),(1,0,0,0),(1,0,0,1),(1,1,0,1)\}$, we have
\begin{equation*}
\left\{
\begin{array}{l}
 s_i=\frac{q}{2}\sum\limits_{l=1}^{m-1}y_{{\pi(l)}}y_{{\pi(l+1)}}-\sum\limits_{l=1}^{m}v_ly_l-v_0+(\frac{q}{2}m+\frac{q}{2}+\frac{q}{2}y_{\pi(m)}-\sum\limits_{l=1}^{n}v_l),\\
 t_i=s_i+\frac{q}{2}y_{\pi(1)}.
\end{array}
\right.
\end{equation*}
\end{itemize}
      Hence, we know that
      $\mathbf{S}_0=\mathbf{S}_2=\mathbf{S}_3=\mathbf{S}_5=\mathbf{S}_7$, $\mathbf{S}_1=\mathbf{S}_{10}=\mathbf{S}_{11}$, $\mathbf{S}_4=\mathbf{S}_{12}$, $\mathbf{S}_{6}=\mathbf{S}_{8}=\mathbf{S}_{9}=\mathbf{S}_{13}$.
      Then, by Lemma \ref{L_Yu}, $(\mathbf{S}^{N},\mathbf{T}^{N})$ is a $(14N,12N)-\mathrm{ZCP}$.

Following similar arguments as provided in Case 2, we can obtain
\begin{eqnarray*}
\xi^{s_{{i},{h}}-s_{i,g}}+\xi^{s_{{i},{h'}}-s_{i,g'}}+\xi^{t_{{i},{h}}-t_{i,g}}+\xi^{t_{{i},{h'}}-t_{i,g'}}=0,
\end{eqnarray*}
where $i=0,1,\cdots,13$.

Similarly, $L_1=14\cdot 2^{n}$, we have the same result.
\end{itemize}
\end{proof}
  \end{Theorem}
\begin{Example}\label{E_4}
Taking $q=2$, $m=2$, $n=0$, $L_1=14$, $L_2=4$, $\pi=(1,2)$, $v_0=v_1=v_2=0$.
The 2-D generalized Boolean function is $f=y_1y_2+y_2(x_1+x_4+x_1x_2+x_2x_3+x_2x_4+x_3x_4 +x_1x_4+x_1x_2x_3+x_1x_3x_4)+(x_1+x_2+x_1x_2+x_1x_3+x_2x_4+x_1x_2x_4)$.
The array pair
$(\mathcal{s},\mathcal{t})=(\mathcal{f},\mathcal{f}+\mathcal{y}_1)$ forms a $((14,4),(12,4))-\mathrm{ZCAP}$ of which
\begin{eqnarray*}
&&\mathcal{s}^{T}=\left(
\begin{array}{ccccccccccccccccccccccccccccccccc}
     0&     0&     0&     0&     1&     0&     1&     0&     1&     1&     0&     0&     1&     1\\
     0&     1&     0&     0&     1&     0&     0&     0&     0&     0&     1&     1&     1&     0\\
     0&     0&     0&     0&     1&     0&     1&     0&     1&     1&     0&     0&     1&     1\\
     1&     0&     1&     1&     0&     1&     1&     1&     1&     1&     0&     0&     0&     1\\
\end{array}
\right),\\
&&\mathcal{t}^{T}=\left(
\begin{array}{ccccccccccccccccccccccccccccccccc}
     0&     0&     0&     0&     1&     0&     1&     0&     1&     1&     0&     0&     1&     1\\
     0&     1&     0&     0&     1&     0&     0&     0&     0&     0&     1&     1&     1&     0\\
     1&     1&     1&     1&     0&     1&     0&     1&     0&     0&     1&     1&     0&     0\\
     0&     1&     0&     0&     1&     0&     0&     0&     0&     0&     1&     1&     1&     0\\
\end{array}
\right),
\end{eqnarray*}
where $\mathcal{s}^{T}$ and $\mathcal{t}^{T}$ are the transpose of $\mathcal{s}$ and $\mathcal{t}$. The aperiodic autocorrelation sum is depicted in Fig.2.
\begin{figure}[htb]\label{Tu_2}
\centering
\includegraphics[width=8cm]{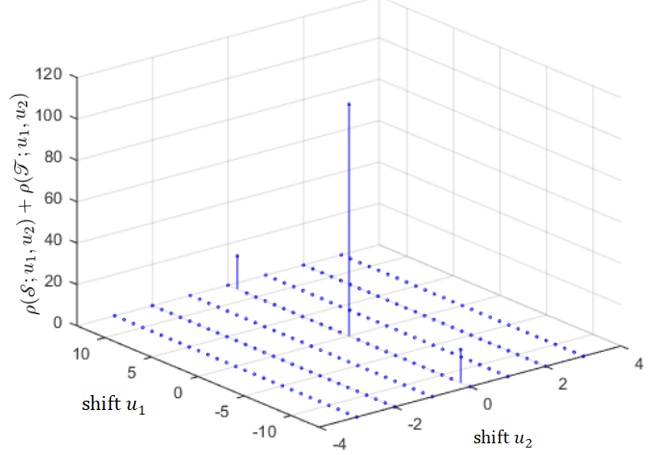}
\caption{Sum of autocorrelations of the $\mathrm{ZCAP}$ $(\mathcal{S},\mathcal{T})$ in Example \ref{E_4}.}
\end{figure}
\end{Example}
\begin{Example}\label{E_5}
Taking $q=2$, $m=2$, $n=1$, $L_1=28$, $L_2=2$, $\pi=(1,2)$, $v_0=v_1=v_2=0$.
The 2-D generalized Boolean function is $f=y_1y_2+y_2(x_1+x_4+x_1x_2+x_2x_3+x_2x_4+x_3x_4 +x_1x_4+x_1x_2x_3+x_1x_3x_4)+(x_1+x_2+x_1x_2+x_1x_3+x_2x_4+x_1x_2x_4)$.
The array pair
$(\mathcal{s},\mathcal{t})=(\mathcal{f},\mathcal{f}+\mathcal{y}_1)$ forms a $((28,2),(24,2))-\mathrm{ZCAP}$ of which
\begin{eqnarray*}
&&\mathcal{s}^{T}=\left(
\begin{array}{cccccccccccccccccccccccccccccccccccccccccccccccccccccccccccccccccc}
0&0&0&0&0&0&0&0&1&1&0&0&1&1&0&0&1&1&1&1&0&0&0&0&1&1&1&1\\
0&1&1&0&0&1&0&1&1&0&0&1&0&1&0&1&0&1&0&1&1&0&1&0&1&0&0&1\\
\end{array}
\right),\\
&&\mathcal{t}^{T}=\left(
\begin{array}{cccccccccccccccccccccccccccccccccccccccccccccccccccccccccccccccccc}
0&1&0&1&0&1&0&1&1&0&0&1&1&0&0&1&1&0&1&0&0&1&0&1&1&0&1&0\\
 0&0&1&1&0&0&0&0&1&1&0&0&0&0&0&0&0&0&0&0&1&1&1&1&1&1&0&0\\
\end{array}
\right),
\end{eqnarray*}
where $\mathcal{s}^{T}$ and $\mathcal{t}^{T}$ are the transpose of $\mathcal{s}$ and $\mathcal{t}$. The aperiodic autocorrelation sum is depicted in Fig.3.
\begin{figure}[htb]\label{Tu_3}
\centering
\includegraphics[width=8cm]{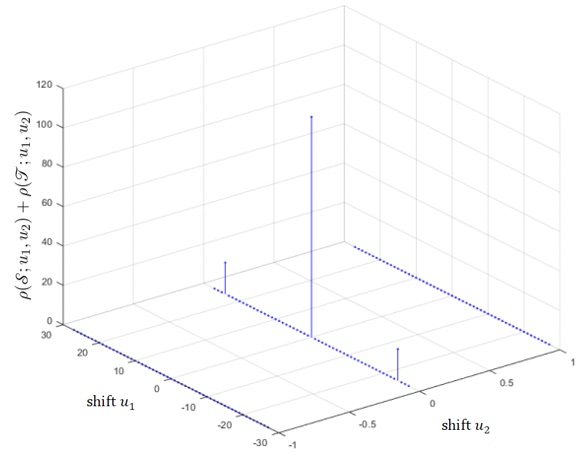}
\caption{Sum of autocorrelations of the $\mathrm{ZCAP}$ $(\mathcal{S},\mathcal{T})$ in Example \ref{E_5}.}
\end{figure}
\end{Example}
\section{Conclusions}\label{Sec-Conclusions}
In this paper, construction methods for designing complementary array pairs have been explored. The derived constructions follow two approaches; the first one can be viewed as a secondary-like construction, and the second one is a direct construction. Specifically, Theorem 1 presents a new construction of 2-D ZCAPs based on 1-D ZCPs, and Theorem 2 offers a direct construction of 2-D ZCAPs based on generalized Boolean functions. Besides, we point out that the construction derived from Corollary 1 covers previous constructions presented in \cite{2020-pai} and \cite{2021-4-pai}. We emphasize that, compared to \cite{2021-4-pai} and \cite{2021-Roy}, our proposed 2-D ZCAPs have the largest 2-D $\mathrm{ZCZ}_{\mathrm{ratio}}=\frac{6}{7}$ based on 2-D generalized Boolean functions. In addition, compared to \cite{2019-pai} and \cite{2021-pai}, our proposed 2-D ZCAPs also have the largest 2-D $\mathrm{ZCZ}_{\mathrm{ratio}}$ based on 1-D ZCPs.

%


\begin{thebibliography}{99}
\bibitem{2018-Adhikary}A. R. Adhikary, S. Majhi, Z. Liu, and Y. L. Guan, ``New sets of even-length binary Z-complementary pairs with asymptotic ZCZ ratio of 3/4,'' {\it IEEE Signal Process. Lett.}, vol. 25, no. 7, pp. 970-973, Jul. 2018.

\bibitem{2003-7-Borwein} P. B. Borwein and R. A. Ferguson, ``A complete description of Golay pairs for lengths up to 100,''  {\it Math. Comput.}, vol. 73, no. 246, pp. 967-985, 2004.

\bibitem{2017-Chen} C.-Y. Chen, ``A novel construction of Z-complementary pairs based on generalized Boolean functions,'' {\it IEEE Signal Process. Lett.}, vol. 24, no. 7, pp. 987-990, Jul. 2017.

\bibitem{2002-Craigena} R. Craigen, W. Holzmann and H. Kharaghani, ``Complex Golay sequences: structure and applications,'' {\it Discrete Mathematics}, vol. 252, no. 1-3, pp. 73-89, May 2002.

\bibitem{1999-GDJ} J. A. Davis and J. Jedwab, ``Peak-to-mean power control in OFDM, Golay complementary sequences, and Reed-Muller codes,'' {\it IEEE Trans. Inf. Theory}, vol. 45, no. 7, pp. 2397-2417, Nov. 1999.

\bibitem{2003-Farka?} P. Farka\v{s} and M. Turcs\'{a}ny, ``Two-dimensional orthogonal complete complementary codes,'' {\it in Proc. Joint IST Workshop on Mobile Future and Symp. on Trends in Commun.}, Bratislava, Slovakia, pp. 21-24, Oct. 2003.

\bibitem{2007-8-Fan} P. Fan, W. Yuan, and Y. Tu, ``Z-complementary binary sequences,'' {\it IEEE Signal Process. Lett.}, vol. 14, no. 8, pp. 509-512, Aug. 2007.

\bibitem{1951-Golay} M. J. E. Golay, ``Static multislit spectrometry and its application to the panoramic display of infrared spectra,'' {\it J. Opt. Soc. Am.}, vol.41, no.7, pp. 468-472, Jul. 1951.

\bibitem{1982-Golomb} S. W. Golomb and H. Taylor, ``Two-dimensional synchronization patterns for minimum ambiguity,'' {\it IEEE Trans. Inf. Theory}, vol. 28, no. 4, pp. 600-604, Jul. 1982.

\bibitem{2019-Gu}Z. Gu, Y. Yang, and Z. Zhou, ``New sets of even-length binary Z-complementary pairs,'' {\it in Proc. 9th IEEE Int. Workshop Signal Des. Appl. Commun.}, pp. 1-5, Oct. 2019.


\bibitem{1983-Hershey} J. E. Hershey and R. Yarlagadda, ``Two-dimensional synchronisation,'' {\it Electronics Letters}, vol. 19, no. 19, pp. 801-803, Sep. 1983.



\bibitem{2018-Kumari} P. Kumari, J. Choi, N. Gonzalez-Prelcic, and R. W. Heath, ``IEEE 802.11ad-based radar: An approach to joint vehicular communication radar system,'' {\it IEEE Trans. Veh. Technol.}, vol. 67, no. 4, pp. 3012-3027,  Apr. 2018.

\bibitem{2011-1-Li} X. Li, P. Fan, X. Tang, and Y. Tu, ``Existence of binary Z-complementary pairs,'' {\it IEEE Signal Process. Lett.}, vol. 18, no. 1, pp. 63-66, Jan. 2011.

\bibitem{2011-Li}Y. Li and C. Xu, ``Construction of two-dimensional periodic complementary array set with zero-correlation zone,'' {\it in Proc. Int. Workshop on Signal Design and Its Appl. in Commun.}, Guilin, China, pp. 104-107,  Oct. 2011.

\bibitem{2014-Liu-even} Z. Liu, U. Parampalli, and Y. L. Guan, ``On even-period binary Z-complementary pairs with large ZCZs,'' {\it IEEE Signal Process. Lett.}, vol. 21, no. 3, pp. 284-287, Mar. 2014.

\bibitem{2014-odd} Z. Liu, U. Parampalli, and Y. L. Guan, ``Optimal odd-length binary Z-complementary pairs,'' {\it IEEE Trans. Inf. Theory}, vol. 60, no. 9, pp. 5768-5781, Sep. 2014.


\bibitem{2020-pai}C.-Y. Pai and C.-Y. Chen, ``Constructions of two-dimensional Golay complementary array pairs based on generalized Boolean functions,'' {\it in Proc. IEEE Int. Symp. Inf. Theory}, pp. 2931-2935, Jun. 2020.

\bibitem{2021-4-pai} C.-Y. Pai, C.-Y. Chen, ``A novel construction of Two-Dimensional Z-Complementary Array Pairs with Large Zero Correlation Zone,'' {\it IEEE Signal Process. Lett.}, vol. 28, pp. 1245 -1249, 2021.


\bibitem{2021-pai}C.-Y. Pai, Y.-T. Ni, and C.-Y. Chen, ``Two-dimensional binary Z-complementary array pairs,'' {\it IEEE Trans. Inf. Theory}, vol. 67, no. 6, pp. 3892-3904, Jun. 2021.

\bibitem{2019-pai}C.-Y. Pai, Y.-T. Ni, Y.-C. Liu, M.-H. Kuo, and C.-Y. Chen, ``Constructions of two-dimensional binary Z-Complementary array pairs,'' {\it in Proc. IEEE Int. Symp. Inf. Theory (ISIT)}, pp. 2264-2268, Jul. 2019.

\bibitem{2021-Roy}A. Roy, P. Sarkar, and S. Majhi, ``A direct construction of q-ary 2-D Z-complementary array pair based on generalized Boolean functions,'' {\it IEEE Commun. Lett.}, vol. 25, no. 3, pp. 706-710, Mar. 2021.


\bibitem{2019-Shen}B. Shen, Y. Yang, Z. Zhou, P. Fan, and Y. L. Guan, ``New optimal binary Z-complementary pairs of odd length $2^m+3$,'' {\it IEEE Signal Process. Lett.}, vol. 26, no. 12, pp. 1931-1934, Dec. 2019.

\bibitem{2001-Spasojevic} P. Spasojevic and C. N. Georghiades, ``Complementary sequences for ISI channel estimation,'' {\it IEEE Trans. Inf. Theory}, vol. 47, no. 3, pp. 1145-1152, Mar. 2001.


\bibitem{2004-Turcs} M. Turcs\'{a}ny and P. Farka\v{s}, ``New 2D-MC-DS-SS-CDMA techniques based on two-dimensional orthogonal complete complementary codes,'' {\it in Proc. Multi-Carrier Spread-Spectrum}, Dordrecht, Netherlands, pp. 49-56, Jan. 2004.



 \bibitem{2014-Wang} Z. Wang, M. G. Parker, G. Gong, and G. Wu, ``On the PMEPR of binary Golay sequences of length $2^n$,'' {\it IEEE Trans. Inf. Theory}, vol. 60, no. 4, pp. 2391-2398, Apr. 2014.


\bibitem{1983-Weathers} G. Weathers and E. M. Holliday, ``Group-complementary array coding for radar clutter rejection,'' {\it IEEE Trans. Aerospace and Electronic Systems}, vol. AES-19, no. 3, pp. 369-379, May 1983.

\bibitem{2018-Xie}C. Xie and Y. Sun, ``Constructions of even-period binary Z-complementary pairs with large ZCZs,'' {\it IEEE Signal Process. Lett.}, vol. 25, no. 8, pp. 1141-1145, Aug. 2018.


\bibitem{2021-YU}T. Yu, X. Du, L. Li, and Y. Yang, ``Constructions of Even-Length Z-Complementary Pairs With Large Zero Correlation Zones,'' {\it IEEE Signal Process. Lett.}, vol. 28, pp. 828-831, Jan. 2021.

\bibitem{2004-5-zeng}F. Zeng, Z. Zhang, and L. Ge, ``Theoretical limit on two dimensional generalized complementary orthogonal sequence set with zero correlation zone in ultra wideband communications,'' {\it in Proc. IEEE UWBST${\&}$IWUWBS}, Kyoto, Japan, pp. 197-201, May 2004.

%

\end{thebibliography}
\end{document}